# Bicycle cycles and mobility patterns

## Exploring and characterizing data from a community bicycle program

Andreas Kaltenbrunner
andreas.kaltenbrunner@barcelonamedia.org

Rodrigo Meza
rodrigo.meza@barcelonamedia.org

Jens Grivolla
jens.grivolla@barcelonamedia.org

Joan Codina
joan.codina@barcelonamedia.org

Rafael Banchs
rafael.banchs@barcelonamedia.org

Barcelona Media Centre d'Innovació,
Ocata 1,
08003 Barcelona, Spain

## ABSTRACT

This paper provides an analysis of human mobility data in an urban area using the amount of available bikes in the stations of the community bicycle program Bicing in Barcelona. The data was obtained by periodic mining of a KML-file accessible through the Bicing website. Although in principle very noisy, after some preprocessing and filtering steps the data allows to detect temporal patterns in mobility as well as identify residential, university, business and leisure areas of the city. The results lead to a proposal for an improvement of the bicing website, including a prediction of the number of available bikes in a certain station within the next minutes/hours. Furthermore a model for identifying the most probable routes between stations is briefly sketched.

## 1. INTRODUCTION

Human mobility patterns have received a certain amount of attention in recent studies. However, it is not a straightforward task to obtain data which allows a large scale study, mostly due to privacy issues. Notable exceptions where the authors were able to overcome those difficulties include the use of geotagged photos [3] and location data of mobile phones [9, 4], or analyzing the circulation of individual banknotes [2] and civil aviation traffic [7] to reconstruct geospatial data of human displacements in different distance-scales.

Some of these studies focus on the trajectories of individuals which are reconstructed in several different manners. Large distance displacements can be deduced from aviation traffic and have then been applied to predict the spread of infectious diseases [7]. Another quite ingenious way of the same authors to interfere middle and large scale trajectories was analyzing the circulation data of banknotes provided by individual users at an online bill-tracking system [2]. This study showed that human travel distances can be described by a two-parameter continuous-time random walk model.

Shorter distances have been analyzed in great detail in [4], using position data of individual mobile users. The authors showed that individuals follow simple and reproducible patterns of mobility in their everyday displacements, a fact that has not been found in [2] for middle and large scale trajectories. Another type of short distance patterns have been analyzed in [3], where the focus changed from everyday life patterns to the behavior of tourists in foreign cities. Their spatio-temporal data was deduced from geo-referenced photos and the obtained results where contrasted with mobile phone usage.

A case where, on the contrary to the above described studies, only aggregate spatio temporal data is available (e. g. the number of persons at time $x$ in place $y$), which does not allow the identification of individual trajectories, was analyzed in a recent study [9]. Data of aggregate mobile phone usage allowed to construct activity cycles for different locations, with clear differences between working day and weekend patterns as well as a characterization of certain areas within the city by a cluster analysis.

Here we perform a similar study using a different type of aggregate data to infer human mobility patterns. The input spatio temporal data, which has been obtained by a web mining process, is the number of bicycles in the approximately 400 different stations of Barcelona's community bicycle program Bicing. To our knowledge this is the first study using this type of mobility data.

The aims of this study are twofold. First, we want to obtain a description of the general patterns and activity cycles, which can be deduced from this type of data and second, we want to check if knowledge of those patterns can lead to a prediction of future behavior, which would allow to improve the current web-service of bicing and in turn increase users satisfaction with the system. Knowledge of those patterns could lead to an optimization of the bicing system itself, allowing the operator to predict shortage or overflow of bicycles in certain stations well in advance and adapt its redistribution schedule accordingly on the fly.

Prediction of Bicing activity is a problem related to traffic congestion control, which has been analyzed traditionally for vehicular traffic. See for example [6] for a review on this subject. Related problems have been investigated also in the context of web-server traffic congestion. A recent study [1] used linear fits of activity to predict web traffic hot-spots. Here we use a technique based on activity cycles more related to [8] where different patterns reflecting a websites activity cycle where used to predict the number of comments a news-item would receive.

To bridge the gap between studies using individual or aggregate displacement data we also briefly sketch a maximum entropy [5] based model which could allow the detection of probable trajectories out of the aggregate data. Such problems have also been extensively studied in the context of vehicular traffic flow [6].

The rest of paper is organized as follows. We first give a more detailed description of the Bicing system in section 1.1. Afterwards we describe details of the data retrieval (section 1.2) and basic quantities of the collected data (section 1.3). In the results part of the article we first describe the patterns of activity in some stations in section 2 and then take a global picture analyzing the activity cycle of the entire city measured by the amount of bicycles in the stations (section 2.2) and their variation as spatial distribution (section 2.3). Then in section 3 some clustering is performed and in section 4 we apply the findings to predict future activity. Finally, we present the model for reconstructing probable trajectories in section 5 and the conclusions in section 6.

## 1.1 Bicing

Bicing is an urban community bicycle program, managed and maintained in partnership by the city council of Barcelona and the *Clear Channel Communications* Corporation. Bicing is mainly oriented to cover small and medium daily routes of users within the Barcelona city area.

Users register into the system paying a fixed amount for a yearly subscription and receive an RFID Card that allows them an unlimited usage through the year, where the first half hour of usage is free and subsequent half hour intervals are charged at 0.30 euros up to a maximum of 2 hours. Exceeding this period is penalized with 3 euros per hour. There are approximately 400 stations distributed all through the city, where each station has a fixed number of slots, either empty (without a bicycle), occupied (holding a bicycle) or out of service, either because the slot itself or the bicycle it contains is marked as damaged. Whenever a subscriber needs to use a bicycle, he must select one from a station with occupied slots, travel to his destiny station, and leave it there on a free slot. The system registers every time a user takes or parks a bike in a slot. Bicycles can be withdrawn from the stations from Monday to Friday between 5:00 and 24:00. On Saturday and Sunday the service is open 24h. Outside of these time windows the bicycles can only be returned but not withdrawn.

There are two cases in which the system does not allow a user to fulfill his route:

1. The origin station does not have any available bicycles.
2. The destiny station does not have any empty slots to park in.

When any of these situations occur, users needing a free slot or a bicycle have to choose between waiting at the station, going to another station or take other means of transportation. In order to reduce these type of situations, there are trucks which move bicycles from highly loaded stations to empty ones. However, in practice users do not wait for these trucks since they do not have a fixed schedule nor ensure a maximum response time to fix problems at a station.

To allow users to plan their routes in advance, the bicing system provides on their website a map of stations[1], where users can check the status of the stations (amount of available bikes and empty slots) close to their departure and arrival points. However, this information is only available at the specific moment when the user queries the system. The service does not provide a history of previous loads to the stations[2] or an expected load of the destiny station at the time that the user gets to it.

## 1.2 Data retrieval

The Bicing website provides an information service for users through the Google maps API. It shows a map of Barcelona overlayed with small markers indicating station positions and the amount of available bicycles and free slots for every station. Data is inserted into the map using JavaScript code with a string variable that contains a KML geospatial annotation document. This KML document defines the next information for each station:

1. station name
2. graphic icon to be inserted in the map
3. latitude and longitude
4. number of available bicycles
5. number of free slots

In order to analyze the dynamics of station loads, we have been collecting these KML documents since May 15th every two minutes, parsing it and storing in a MySQL database all the relevant information, such as the station name, localization, available bicycles and free slots. As the Bicing network changes from time to time, new stations are added automatically to the database when they first appear in the KML files collected from the bicing website.

[1] `www.bicing.com/localizaciones/localizaciones.php`

[2] A nice personal project (`http://statistings.com`) improves the service by providing the daily progression of the number of bicycles in the stations.

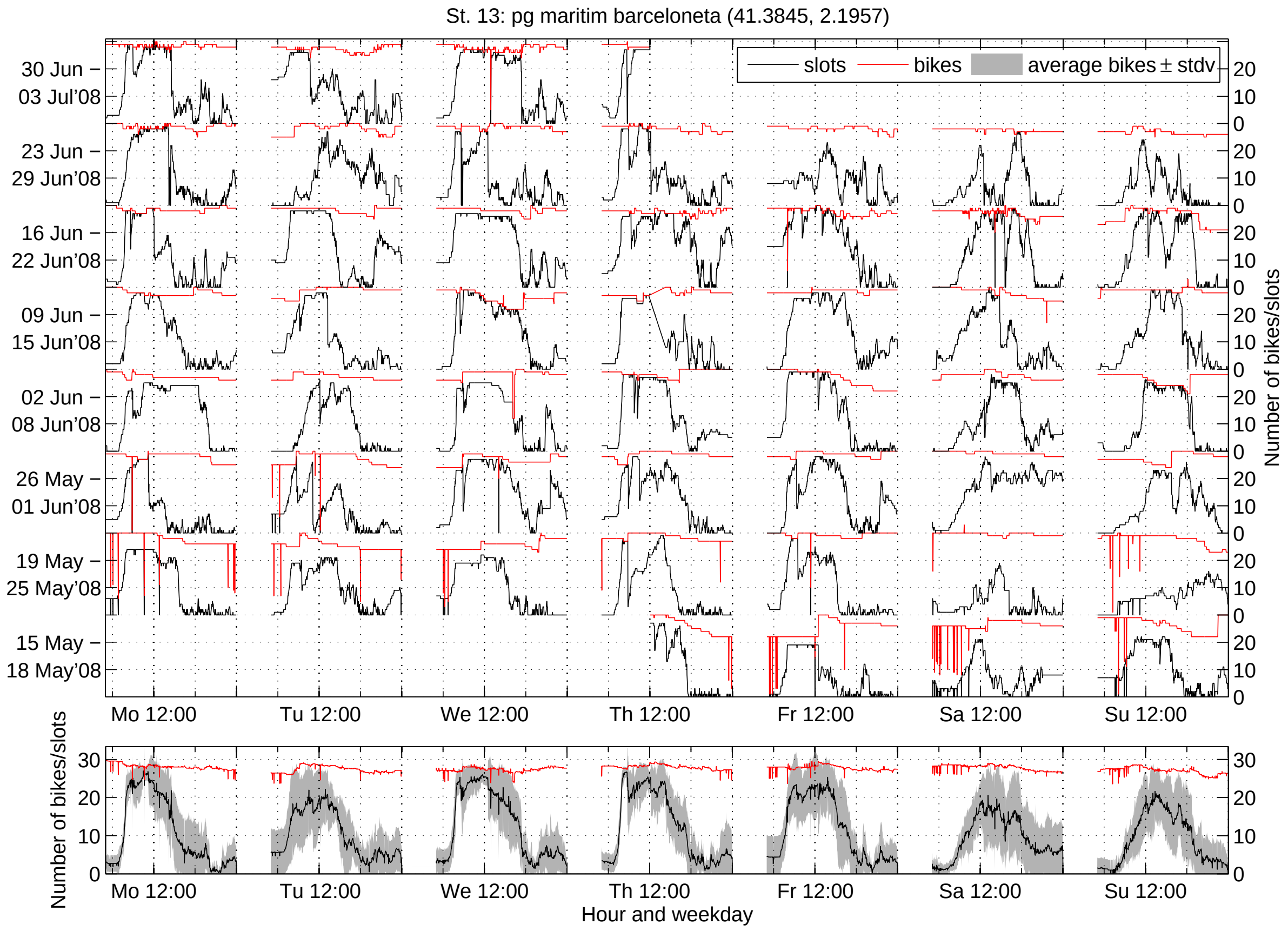

**Figure 1: Sequence of the number of bicycles in the station (black line), and the total number of slots (red line) of an example station next to the beach. Bottom row shows the average weekly pattern of this station. Gray areas correspond to mean ± one stdv.**

### 1.3 Basic quantities of the data collected

Due to a problem in the bicing web-service, data after the 3rd of June was updated only once or twice a day and could not be used for our study. We base our results therefore on the data recollected during the 7 weeks between 12:00, May 15th and 12:00, July 3rd, 2008. We also initially did not collect data during Bicing's closing hours on weekdays between 0:01 and 5:00, which restricts our analysis further to the time-window between 5:00 and 24:00.

In total, we collected data from 377 stations with a total of approximate 8700 free slots (three stations, which never contained any bicycles, were omitted from the analysis). The number of slots per station varies between 15 and 39 and the maximum amount of bicycles in the stations observed in our data was 3657. Table 1 summarizes these numbers.

| | |
|---|---|
| number of stations | 377 (374 with data) |
| number of slots | ≈ 8700 |
| slots per station | [15 – 39] |
| max. number of bicycles observed | 3657 |

**Table 1: Principal quantities of the data recollected.**

## 2. ACTIVITY CYCLES

The amount of bicycles available at the different stations allows us to infer activity cycles of Barcelona's population.

### 2.1 Local activity cycles

Before we begin calculating activity cycles we take a closer look at the data recovered from Bicing's web-service.

The top plot in Figure 1 shows an example of the recovered time-series data from a station close to the beach, a hospital and some office and university buildings. The recollection started on Thursday, 15-05-2008 (bottom of the subfigure) and subsequent weeks are drawn with an offset towards the top of the figure. The black lines indicate the amount of available bicycles. For control reasons we also draw the sum of bicycles and empty slots (red line), which in case the station were 100% operationally should correspond to the total number of its slots. However, since often some slots or bicycles are marked as defect and cannot be used, the red lines show some fluctuations. Sometimes they experience a sudden drop during short time intervals (e.g. on Saturday, 17-05-2008 morning), probably caused by a sporadic malfunction in Bicing's data collection system.

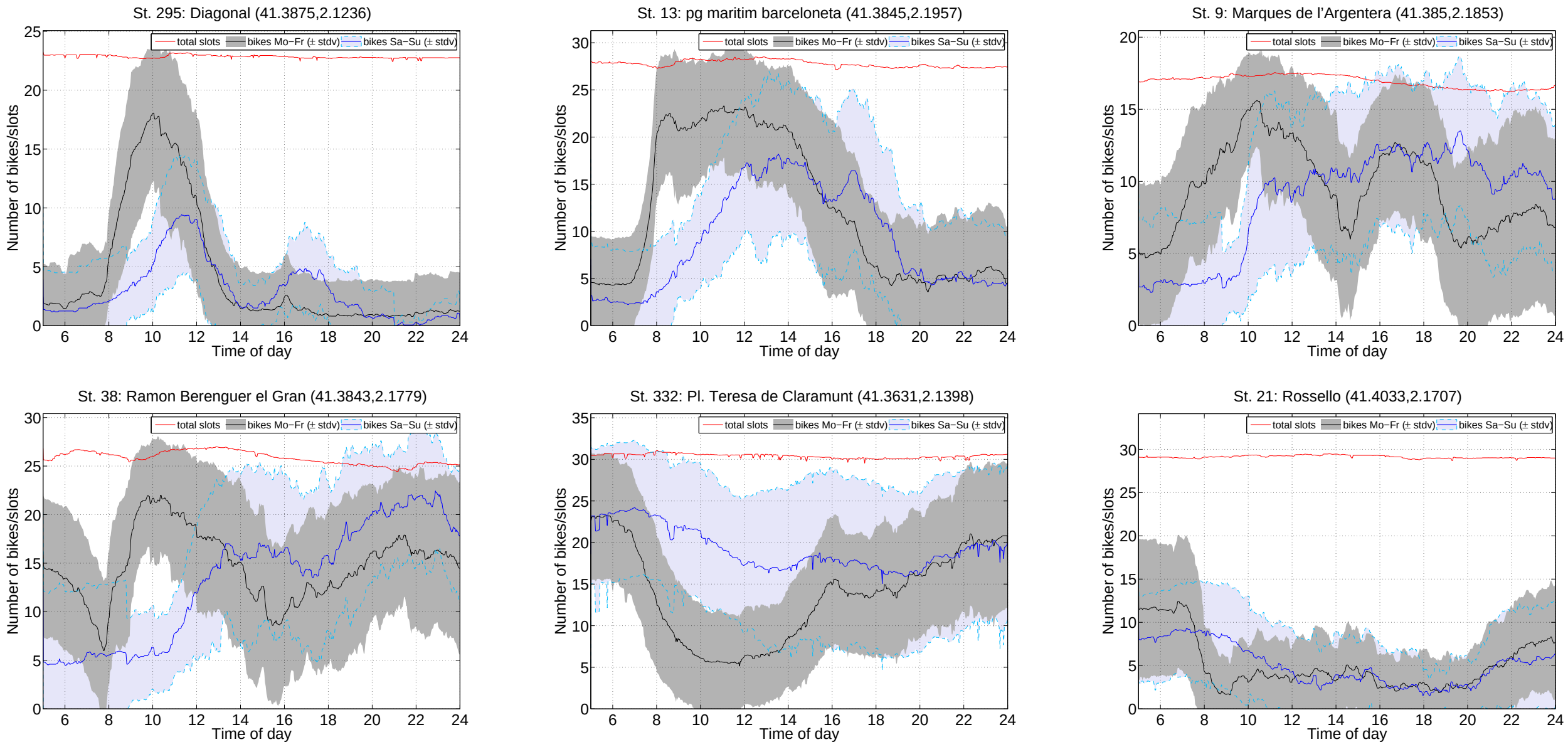

**Figure 2: Average number of available bicycles during working days (black), and weekends (blue lines) for six example stations with different types of activity cycles. Red curve gives the average total number of slots in the station. Gray and blue areas correspond to mean ± one stdv.**

Although the data is quite noisy with some sudden drops in the number of bikes, maybe caused by replacement trucks which move bicycles from occupied stations to empty ones, the mean weekly activity pattern shown in the bottom subplot of Figure 1, allows to average out those fluctuations quite well. We therefore have chosen to ignore those unpredictable truck events in the rest of this study. The relative small standard deviations (black areas) show that the observed patterns are quite stable during the 7 weeks of data we analyzed. Note especially the near zero deviation at the sharp rise in the morning which can be observed from Monday to Friday. The greater standard deviation of the Tuesday pattern is caused by the local holiday on June 24th, whose bicycle pattern is more similar to those of a typical Sunday. We clearly observe two different patterns for weekend and working days.

This is confirmed by a more detailed analysis of these two patterns in Figure 2, where weekend (blue lines) and weekday patterns (black lines) from six different stations are compared. To calculate those patterns we first delete all the elements of the time-series where the total number of slots in the station is below a certain threshold (10). This allows to eliminate most of the moments where we believe the data to be erroneous (e.g. the drops in the red line in Figure 1). We then average those filtered time-series over the days of the corresponding categories and apply a median filter with window length 3 to filter the noise further.

We first focus only on the weekday patterns. The top middle subplot corresponds to the station analyzed in Figure 2 in more detail. We observe very different patterns in the different stations. Station #295 (top left) is close to a university and shows a quite narrow peak in the number of bicycle in the station between 8:00 and 13:00, typical for a university with morning classes only. The following two stations are also close to universities (top middle and right subplots). However, their observed patterns are somehow different. All three stations show the initial rise in activity in the morning. Sharp in station #13 (top middle) and less pronounced in #9 (top right). Station #13 is also close to some important office buildings and a hospital which might explain the sharp raise in activity around 8:00, more prone to a fixed working schedule in companies or hospitals than varying starting hours of university classes. The location close to the beach probably causes the lower decay in the number of bikes in the afternoon hours where beach traffic collides with the leaving students and office and sanitary workers. Station #9 shows more variability. Although more spread than station #295 the morning peak is quite similar. However, this station experiences a second peak starting at 15:00 and reaching its maximum at 16:00 in the afternoon, This might either be caused by people leaving the university to take their lunch elsewhere or a change of shift between morning and afternoon lesson students. Finally, this station also experiences an increase in activity after 20:00 caused with high probability by the popular close-by area of bars and restaurants called "Born".

The station #38 (bottom left) represents another pattern quite different from the previous ones. It shows a drop in activity typical for residential areas where people mainly withdraw bikes to move to their destinations (e.g. station # 21 in the top right corner of Figure 2). However, at 8:00 in the morning the profile of this station changes to a pattern more characteristic for a office/university station. Due to closeness it might serve as backup for nearby university stations which have been run out of free slots to drop off the bikes. It is also situated right in the center between the "Born" and "Barri Gòtic", which explains the increase in the number of bikes in the late evening of people enjoying the nightlife in the city center. Finally, the patterns of stations

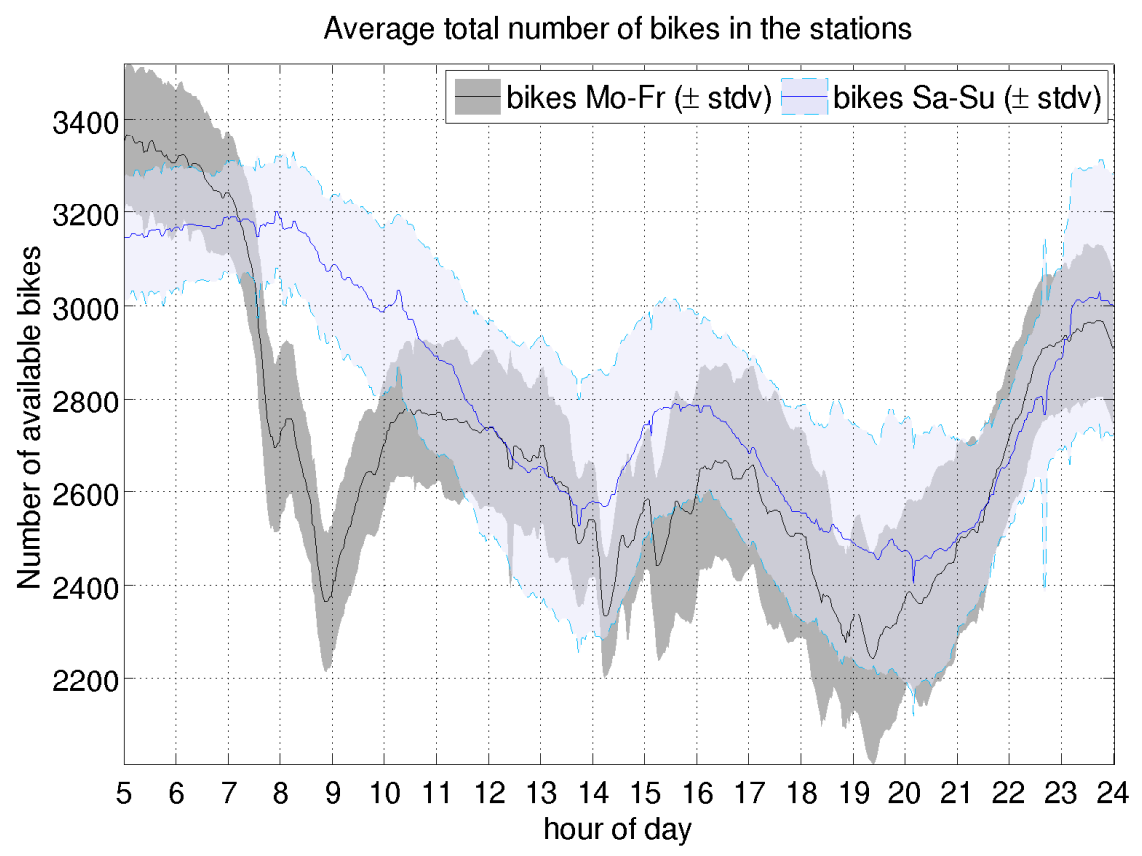

**Figure 3: Average of the total amount of bicycles available in the stations.**

#332 and #21 show opposite cycles to the previous ones, typical for residential areas, where people leave the region during the morning to return later in the afternoon or late evening.

The onsets of activity in the weekend patterns (blue lines) occur later than during working days, or is nearly absent as can be observed for example in station #332 (bottom middle subplot), where only some minor activity is observed. Station #295 shows an interesting bimodal distribution on weekends, which might be caused by a nearby shopping center which attracts afternoon visitors on weekends.

## 2.2 Global activity cycles

If we look instead of the local cycles in the particular stations at the sum of bicycles available at all stations during a certain hour of the day, we get an idea of the global mobility cycle of Barcelona.

In Figure 3 we plot these average cycles of available bicycles for the working days from Monday to Friday (black curve) and the weekend (blue curve). To filter the worst noise out of the data, caused by malfunctions in the system, we use only measurements where the total sum of slots (free and occupied) in all the stations is greater than 8000 and furthermore we apply again a median filter with a window length of 3 to achieve smother curves.

The less bicycles are available for rent in the station the more displacements using them are being performed. First, we analyze the traffic during working days (black line). We observe a first local minimum (i.e. a local maximum in displacements) a little earlier than 8:00, and a second lower one at 9:00. These two minima correspond to the typical starting hours in offices, which in Barcelona varies normally between 8:00 and 10:00. This is further confirmed by the fact that the curve reaches a local maximum at this hour, the time when late starters finally reach their working or study locations. A third lower minimum is observed around 14:00, which might be caused by students who leave their classes. The number of available bicycles increases during people's lunch breaks (typically between 14:00 and 16:00), but when the local maximum at the end of this time span is reached it decays again. Finally, the global minimum number of available bicycles (the maximum in displacements) is reached slightly after 19:00 in the afternoon. Typical finishing time of many working schedules.

The weekend pattern is different in the sense that it does not show the early morning minima. Instead we observe the maximum of available bicycles around 8:00, the equinox between late home-comers from the last parties and early birds starting their day with a bicycle ride. The use of the bikes steadily augments until their number in the stations reaches a local minimum at 14:00 just before lunch time, during which it increases again. Afterwards the number of available bicycles decays again and follows a similar pattern as during working days, although the local maximum at 16:00 occurs slightly earlier and the global minimum slightly later (at 20:00) and is less pronounced than during working days. It is therefore difficult to separate working day from weekend activity only based on afternoon activity, as can be observed as well for most of the stations presented in Figure 2.

Note that initially we only collected data between 5:00 and 24:00, which corresponds to the opening hours of Bicing from Monday to Friday. However, although the users are not allowed to withdraw from a station outside of this time schedule, they can return a bicycle also between 24:00 and 5:00. This explains the difference in the number of bicycles available at the beginning and end of the above described cycle.

The small standard deviations (gray and blue areas in Figure 3) show that the observed cycles are quite stable throughout the period the data was collected. The weekend deviation is slightly greater than its working day counterpart which is caused by the greater number of working days in our data set (35 vs 14) and the more flexible personal time-schedules on weekends.

## 2.3 Mobility patterns

To get a spatial picture of the mobility pattern in the city, we use these local activity cycles together with the stations geo-coordinates (longitude and latitude) and place the difference in the number of bicycles in the stations compared to their amount at 5:00 on the map of Barcelona for different times of the day. Afterwards we interpolate a 3D surface using this difference as color-encoded height[3]. Red stands for a positive difference, i.e. more bikes can be found in this stations than at the beginning of the day, while blue regions show areas whose number of bicycles has been reduced. Green areas indicate a more or less constant relation between incoming and outgoing bicycles. Figure 4 shows such geo-patterns for 6 different hours using the stations working day cycle[4]. At 6:30 (top left subfigure), no big difference form the initial distribution of bicycles in the stations can be observed. At 9:30 however (top right subfigure), just after the morning minimum in Figure 3, we observe quite a different picture. Several areas change color either into deep red or dark blue. Blues regions correspond to mainly residential areas, from which people move out, while the red hot-spots are found mainly close to university and business quarters. Interestingly, although the number of bicycles in the station increases by roughly 400 until 12:00 in Figure 3, the snapshot of the geo-pattern (middle left subplot in Figure 4) at this moment in time does not change very much. The only noticeable difference is that in already red regions the amount of bicycles increases slightly even more. We can conclude that the morning peak in activity leads to quite a narrow band of stations with high bicycle concentration. The band crosses the city starting at its westmost entrance, where one the mayor university area of the city lies, and follows the Diagonal through a business district towards Passeig the Gracia, where it turns right and heads down passing by one of the mayor business and shopping areas and the University of Barcelona to meet the city center and later the sea. There it turns left again to follow the beach towards Port Olympic, leaving out one station in the also mainly residential area of Barceloneta and passing by several campuses of Universitat Pompeu Fabra. Close to Port Olympic we also find important office buildings as well as in a narrow band which grows from there northwards towards Glories. Another area which receives a big surplus in activity is Diagonal Mar, the east-most point of Barcelona, also a region with important business activity and a large shopping center.

[3] Alternatively one can repeat the same procedure with other starting times (e.g. 16:00 to emphasize afternoon patterns).
[4] A similar but simpler spatio temporal visualization by Fabien Girardin using just the evolution of bicycles in the stations during one day can be found at `http://www.girardin.org/fabien/tracing/bicing/`

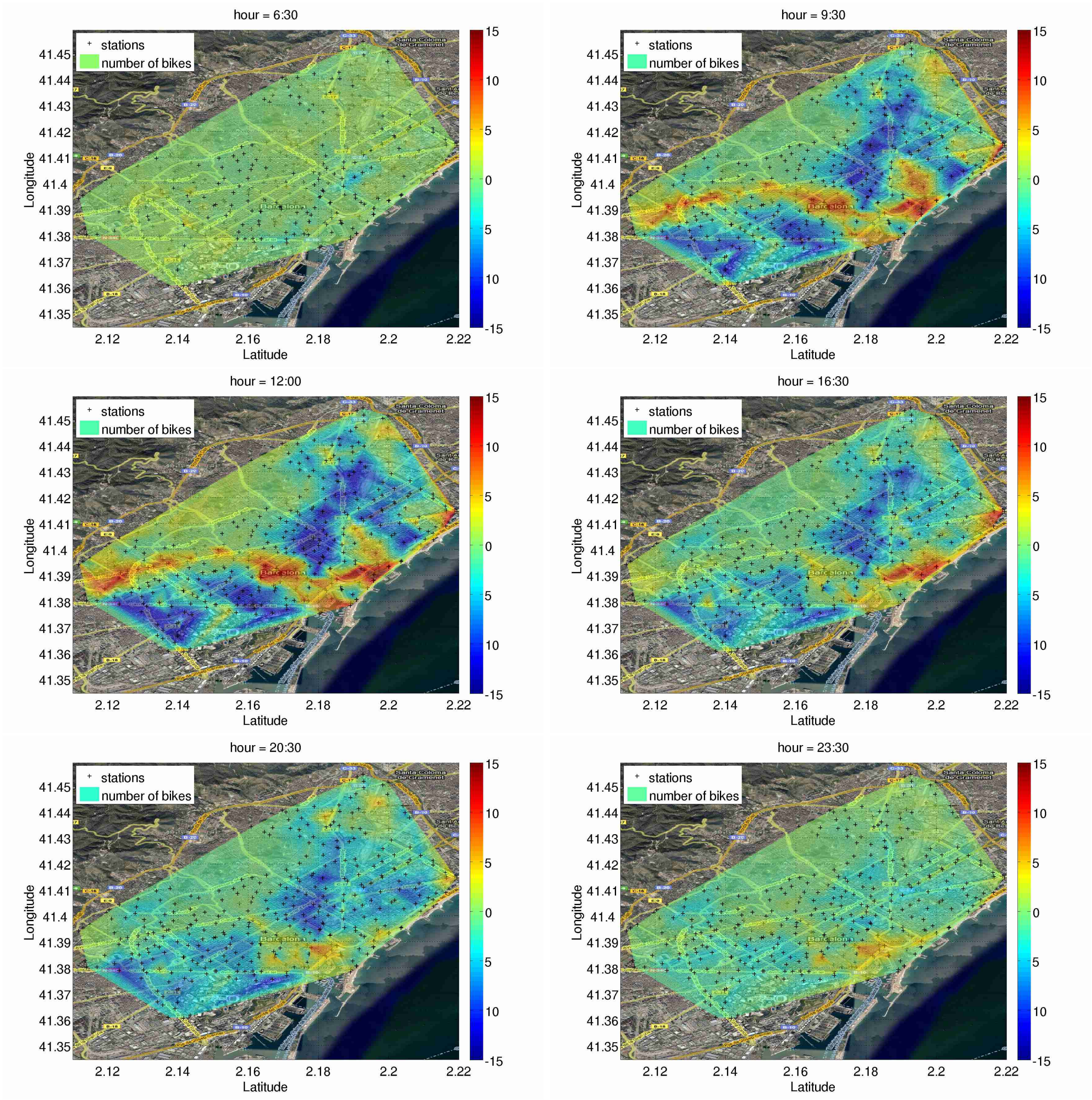

**Figure 4: Geographic mobility patterns: Black crosses indicate the location of bicing station and the color-overlay the average variation during working days in the number of available bikes from the level at 5:00. Blue tones indicate regions which loose bikes while red tones stations which increase their number of bicycles.**

In the afternoon the picture changes, at 16:30 (middle left subfigure) a lot of bicycles have moved away from the previously described hot-spots, and the residential areas get some of their lost bikes back. Only the regions close to Port Olympic remains deeply red, probably now caused mainly by beach traffic. Also Diagonal Mar maintains its bicycles. At 20:30 (bottom left), finally, also those bikes head home again, only some regions in the city center still have a surplus of bicycles, probably caused by people enjoying Barcelona's nightlife. Those regions maintain their bicycles still at 23:30 (bottom right) when most of the remaining stations have recovered all their bikes and green tones. Those stations will recover their bikes during the night.

# 3. CLUSTERING OF ACTIVITY

The human behavior patterns mentioned above suggest that some Bicing stations may show similar cycles depending on the activities occurring around them. To find such similarity patterns, we use the sequences of the average number of bicycles in the stations during working days (as shown for some example stations in 2) and define similarity metrics between those sequences which then allow to generate clusters of stations with similar cycles. We use the following metrics:

**absolute similarity** : Let $p$, $q$ be two bicycle stations and let $T = \{t_i\}_{i=1..n}$ be the set of measure points where station loads are collected, and $s_p(t_i)$ be the average number of available bicycles on station $p$ in the measure point $t_i$. The absolute similarity between stations $p$ and $q$ is defined by:

$$abs_{sim}(p,q) = \frac{1}{\sum_{t_i \in T} |s_p(t_i) - s_q(t_i)|} \quad (1)$$

**relative similarity** Let $p$, $q$, $T$ and $s_p(t_i)$ be defined as before. We define a new function $D_p(t_i)$ as:

$$D_p(t_i) \quad = \quad \begin{cases} 1 & \text{if } s_p(t_{i+1}) - s_p(t_i) \geq 0, \\ -1 & \text{if } s_p(t_{i+1}) - s_p(t_i) < 0, \end{cases} \quad (2)$$

which is basically a slightly modified signum function of the gradient of $s_p$. Then the relative similarity between two stations $p$ and $q$ is defined by:

$$rel_{sim}(p,q) = \sum_{t_i \in T} \frac{1 + D_p(t_i) \times D_q(t_i)}{2} \quad (3)$$

The greater those measures are, the more similar are the involved stations. On one hand, absolute similarity tends to cluster stations according to the exact number of bicycles in every measure point, but does not recognize two stations with the same pattern of use, but a different total number of slots. On the other hand, relative similarity would cluster stations only according to the variations in their usage-pattern, but is not useful to recognize the shape of its cycles.

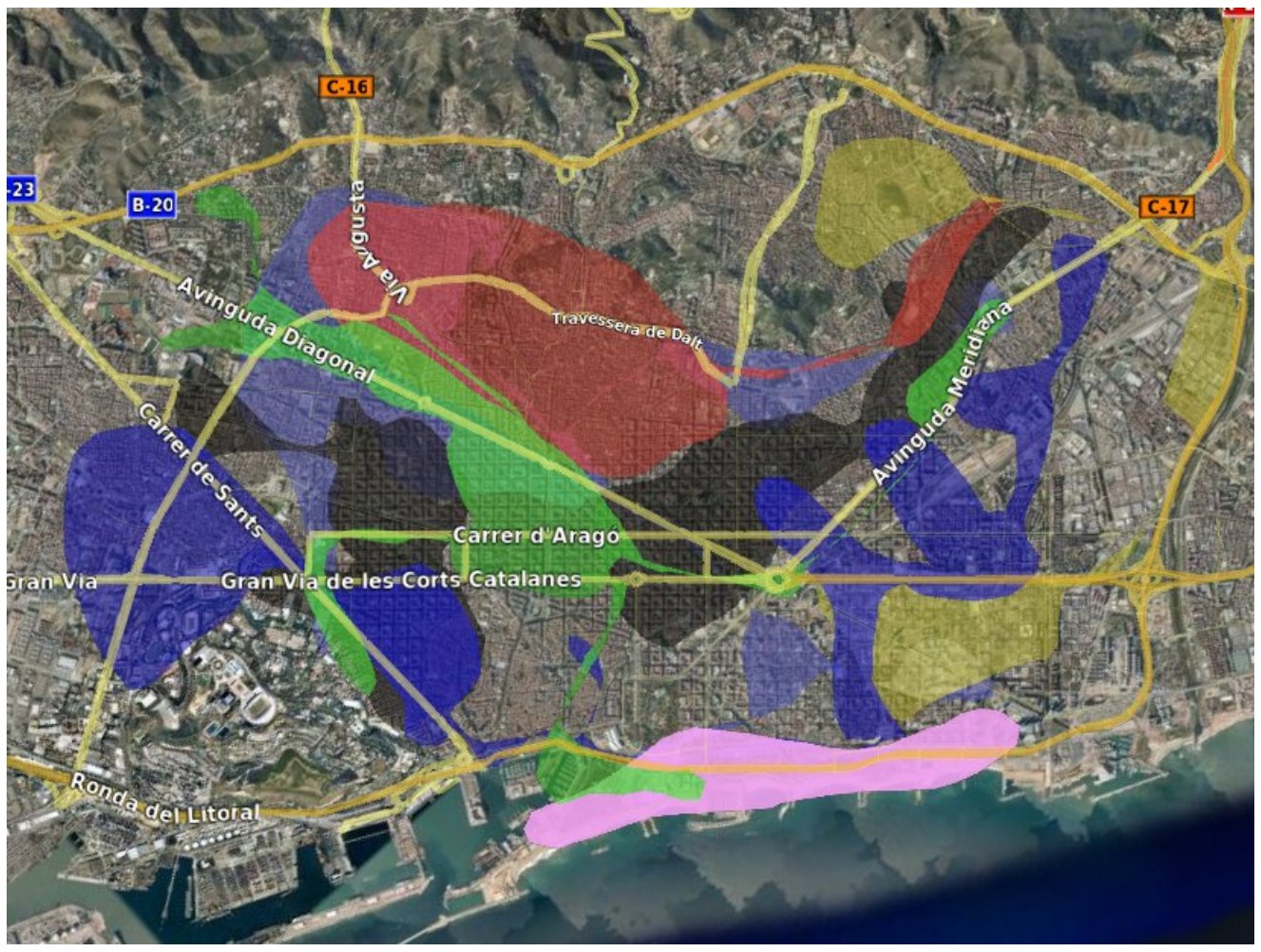

**Figure 5: Clusters grouped by their relative similarity define clear and meaningful zones.**

E.g. an always nearly empty station with little absolute variation in its number of bicycles would be in the same cluster as a station with a large amplitude cycle, if it showed the same occupation pattern variations (receiving bicycles in the mornings, losing them at lunchtime, etc). For this reason, we first calculate absolute similarity between every pair of stations and cluster them using the k-means algorithm. The result is a set of clusters of stations arranged according to their absolute number of bicycles. Afterwards, we use relative similarity to compare the previously generated clusters and using k-means again to generate new clusters of clusters (meta-clusters) with the same usage-pattern, no matter the absolute number of bicycles in them.

To define the optimal number of clusters for the first iteration using absolute similarity, we have calculated internal similarity inside clusters using a variable number of clusters from 2 to the total number of stations. The optimal number of clusters is reached when the minimum internal similarity of all the clusters starts to decrease, that is at 31 clusters.

Once we have obtained this first 31 clusters, we group them using relative similarities between every pair of cluster, getting finally a set of 7 meta-clusters. The geographic zones defined by those meta-clusters are shown in Figure 5 in different colors. The green cluster covers quite well the region with high morning activity detected in Figure 4. Station #295 (Figure 2 top left) is a typical exponent of this cluster. The pink zone corresponds to stations with a typical beach pattern as for example station #13 in Figure 2 (top middle). The black, red and blue meta-clusters cover different types of residential area stations. An example for a blue station can be found in #332 (Figure 2 bottom middle), and in station #21 (bottom right) for a black one. Finally the yellow and light blue meta-clusters show two different peripheric patterns, with a decaying tendency in their number of bicycles during the entire day. This prevalence of outgoing bicycles may require artificial replacements by the operator to ensure a minimum number of bicycles at these stations.

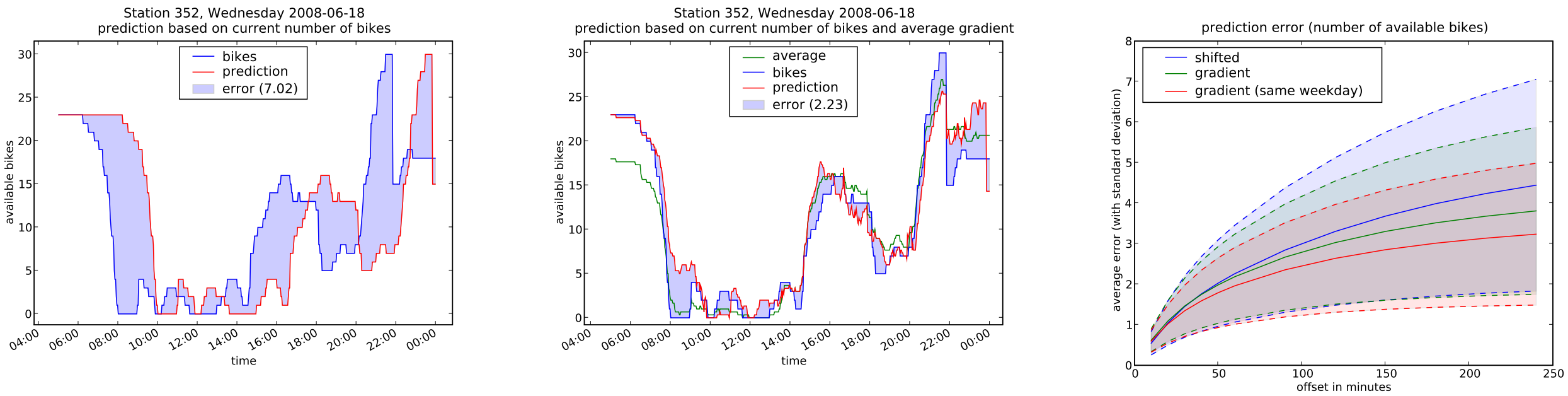

**Figure 6: Prediction of bicycle availability, (left) using only the current value and (middle) adjusted by the average gradient over the other weeks. The right subfigure shows the average error depending on the time offset of the prediction (2 hours for the example day of station #352 shown in the middle and left subfigure).**

Zones in Figure 5 without color overlay correspond to stations whose clusters could not be arranged into meta-clusters. They probably represent a mixture of several different clusters (e.g. the remaining stations #9 and #38 of Figure 2). Future research will try to uncover such combined patterns.

# 4. PREDICTION OF ACTIVITY

In this section, we present initial results on the prediction of bicycles or free slots at a given station at a given time. We compare several simple prediction models, and establish evaluation measures as well as a baseline with which other (more complex) models can be compared.

Our initial set of prediction models is based on the current state of the station as well as aggregate statistics of the station's usage patterns. As the simplest baseline we chose to predict the current state of the station (number of bikes or free slots) for any time in the future. If there are currently 5 available bicycles, the system will predict that in 10 minutes there will still be 5 bicycles available. This corresponds to the best prediction algorithm one can apply using only the present situation as displayed on the actual Bicing website.

The next set of models is based on extrapolating from the current state using the tendencies registered on other dates. To the current number of bikes we add the expected change based on the average gradient in the aggregate model. The aggregate model in this case can be based on all days other than the one for which predictions are made[5], or can be limited to the same day of the week, or split between weekdays and weekends/holidays.

We evaluate the different models by measuring the mean error (difference between predicted and actual availability of bicycles) over all stations and all available dates. This is done for different time offsets, i.e. predicting 10 minutes, 20 minutes, or several hours into the future.

Figure 6 (left) shows an example for the fit obtained using the baseline model (i.e. predicting the current state 2 hours into the future) and (middle) a gradient based prediction (using only data of to the same day of the week) for one particular station and day. The blue curve corresponds to the actual number of bicycles (filtered with a median filter) in the station, while the red one indicates the prediction. In this example we achieve a much lower prediction error (indicated by the light blue areas) using the gradient of the average activity cycle (green curve in Figure 6 middle).

This is confirmed further by Figure 6 (right) where we compare the overall performance of our prediction algorithms as explained above. For very short periods (10 minutes) there is no notable difference between the baseline and other models, which may partly be due to a large number of low activity stations where predicting no change is the safest bet for very short time scales. However, we notice a significantly better performance of prediction algorithms using the activity cycles for larger offsets.

Many enhancements and other approaches remain to be tested, including the incorporation of knowledge about interventions of bicing trucks (by having better data available or detecting it from the available data) and other events that deviate from the "normal" trend. Weather conditions and many other factors may also be taken into account.

# 5. PROBABLE ROUTE IDENTIFICATION

In this section, we will deal with the problem of estimating those routes that are most likely to be transited by the users from the aggregated data that is available to us. Notice that in the actual context of the service provider this problem is not interesting at all, since individual bicycle movements among the stations can be tracked by the system administrator. However, when this information is not available, estimating the most popular routes from the aggregated data is a challenge. Basically, as we will discuss below, this problem is not solvable at all from the observation data alone. But we claim that it is possible to approximate suboptimal solutions for it by means of conditioning the observed data with some additional information, such as the distances among the stations, the average bicycle velocity, and some other common sense implications.

For the appropriate computation of route popularity we should be able to estimate the conditional probability of one bicycle arriving at station $j$ given that it departed from station $i$. In other words, a transition probability matrix $p_{j,i}$ should be estimated for the whole system. As this problem is really involving, we will approach it by considering two important simplifications: first, we will consider the transition probabilities to be time independent of the analysis time interval (we restrict our analysis to the morning behavior from 5:00 to 12:00), and second, we will consider each bicycle to make only one trip during the morning interval. According to this, our problem is described by the following set of equations:

$$F_j = \sum_i p_{j,i} I_i \text{ for } j = 1 \dots N \tag{4}$$

$$\sum_j p_{j,i} = 1 \text{ for } i = 1 \dots N \tag{5}$$

were $N$ is the total number of stations, $I$ is the initial (e.g. at 5:00) distribution of bicycles and $F$ the final (e.g. at 12:00 in our experiments) distribution of bicycles. The first $N$ equations represent the process of aggregation of bicycles arriving from all stations into a final one; and the second set of equations guaranties that the total amount of bicycles is preserved. The system is described by a $2N$ equations but $N^2$ parameters should be determined. So it is strongly undetermined and admits an infinite number of solutions. Even for $N = 2$, the problem remains undetermined since the resulting four equations are linearly dependent.

[5]In a real application setting this would obviously be limited to days prior to the current date.

In our following procedure we use some additional information and some common sense heuristics in order to provide an approximate solution to this problem. It is founded on three important observations:

- Users are more likely to use the service to cover intermediate distances. For short distances (less that 500 meters) the user will rather walk, and for long distances (more than 6 kilometers approximately) the user will rather use another transportation service.
- In the morning interval, most of the users will use the service to move from their home to their working and study area or an alternative transportation system. So during the morning it is expected (and confirmed by the data analyzed above) that some stations mainly serve either as departure or arrival stations.
- Between the groups of departure and arrival stations we can also think about the existence of coupled stations among which a great volume of users should be expected to move. Although this coupling is for sure a many-to-many model, due to the computational expensiveness of this analysis, for simplicity, it will be approximated as a one-to-one sort of phenomenon.

The three previous observations allow us to propose the following maximum-entropy-based model for the transition probabilities we want to approximate:

$$p_{j,i} \approx f_1(dist_{i,j})^{\lambda_1} f_2(sim_{i,j})^{\lambda_2} f_3(coup_{i,j})^{\lambda_3} \tag{6}$$

which is actually a log linear combination of features. The first feature $f_1(dist_{i,j})$ represents a log-normal distribution of the distance between stations. Its parameters have been adjusted to provide a maximum value at 2 kilometers and to be negligible from 7 kilometers. The second feature $f_2(sim_{i,j})$ is a function of the cross-correlation coefficient between the involved stations' average cycle of bicycles (more exactly $f_2(sim_{i,j}) = (1 - xcorr(i,j))/2$), since bicycles are less likely to move between two departure (or arrival) stations.

The third feature, $f_3(coup_{i,j})$, is a ternary function of presumable one-to-one coupled stations. To determine the degree of coupling between two stations we shift the average cycle of observed bicycles at departure stations with respect to the arrival stations by a time factor related to the distance among them (an average speed of 25 km/h was assumed), and we select those pairs of stations that better explained the one-to-one coupling assumption in terms of the total number of bicycles shared by both stations. Every time a departure station $k$ is identified to be coupled with an arrival station $m$, $f_3(coup_{m,k})$ is set to one and the inverse relation $f_3(coup_{k,m}) = 0.1$; for all other cases $f_3(coup_{i,j})$ is set to 0.5. This last assumption helps to assign a probability mass, among other events, to the event of a bicycle remaining in the same station.

The weighting exponents are adjusted to best fit the available aggregated data by means of simplex-based optimization. The columns of the resulting transition probability matrix are normalized after each iteration in order to satisfy condition 5. For the transition probability from 5:00 to 12:00 on weekdays we obtained for our data the weights $\lambda_1 = 0.949$, $\lambda_2 = 1.138$ and $\lambda_3 = 1.116$.

To verify our model we compare in Figure 7 the actual average number of bicycles in the stations at 12:00 (in blue) with the obtained predictions from our model (in green) and a smoothed version of the prediction (in red). The stations or ranked in increasing order by their average number of bicycles. As seen from the figure, the prediction is very noisy, however as illustrated by the smoothed version of it, the tendency is to follow the actual data. This result suggests that, although the prediction accuracy is poor, the assumptions and heuristics used to construct the approximated model can explain the general tendency of the data.

Finally, the most probable routes, according to our proposed model, were extracted from the optimized transition matrix. Figure 8, presents a spatial representation of all bicing stations with the most probable routes among stations depicted. In the figure, those stations identified as departure and arrival stations are presented in blue and red, respectively; while those stations exhibiting different patters are presented in black. The most probable routes (for which transition probabilities were larger than 0.03) are illustrated by the green lines interconnecting the stations. From the picture, several clusters of morning activity can be discovered. It is important to mention that the most probable routes in this map are not necessarily related to heavy traffic of bicycles. They are mainly depicting the relative strength of traffic with respect to the amount of bicycles involved in the stations. This explains the lack of such probable routes in the city center, where most of the activity is to be expected, but this traffic is also is also much more disperse than on the borders of the system.

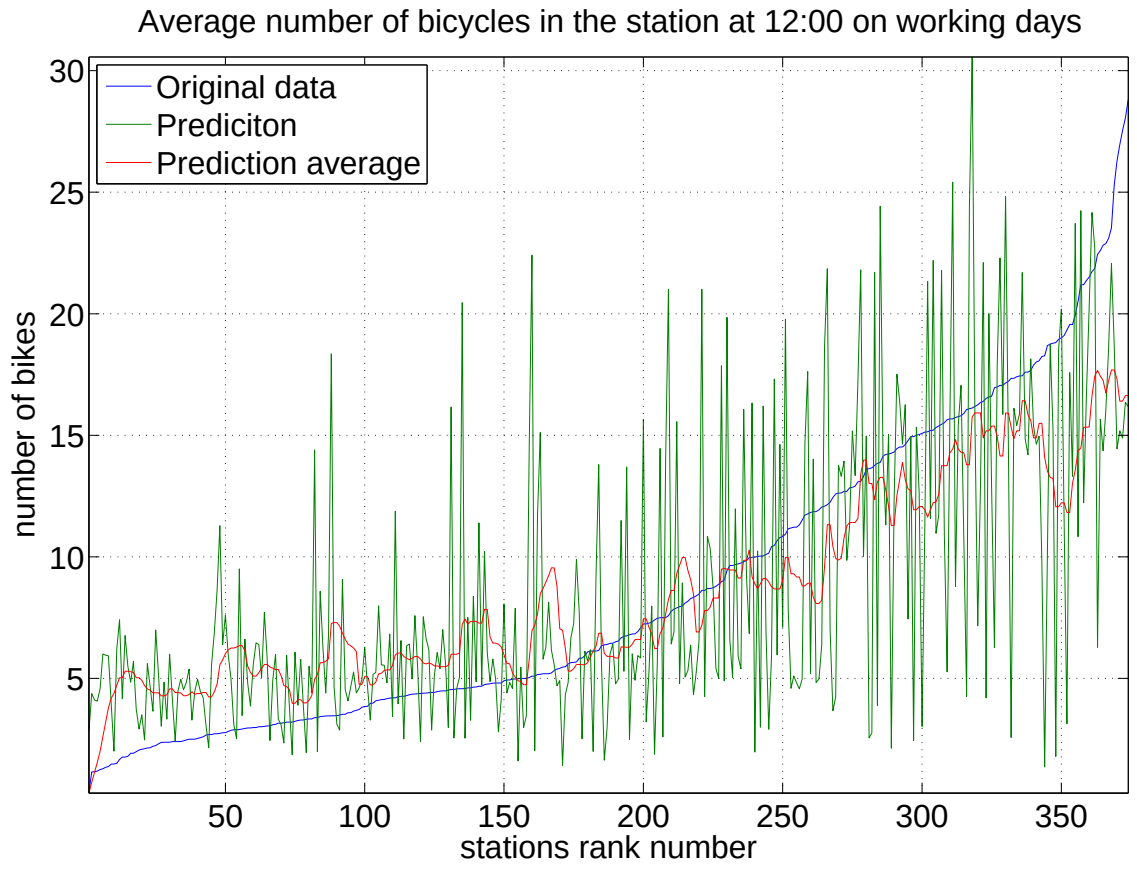

**Figure 7: Actual average bicycle distribution at 12:00 on weekdays (blue), obtained predictions (green) and smoothed predictions (red).**

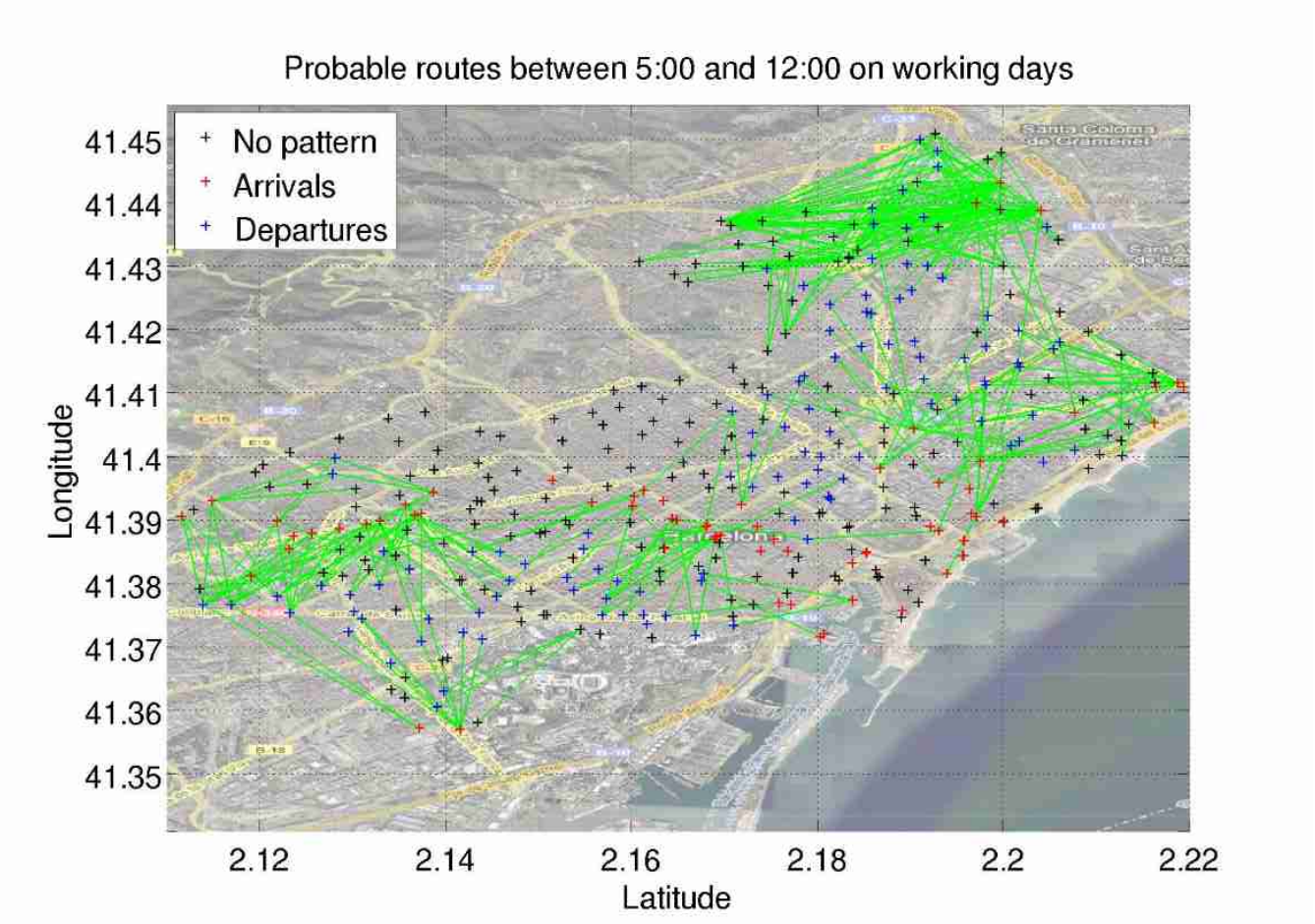

**Figure 8: Most probable routes (green lines) on weekdays between departure (blue), arrival (red) and non-pattern (black) stations.**

# 6. CONCLUSIONS

The approach of mining usage data from community bicycle services presents some advantages against similar studies analyzing cell phone data [9, 4]. The data we use is freely available online for everyone, so one can avoid the typical problems of finding a cell phone company willing to share its data with researchers and the associated privacy and confidentiality issues.

We have shown that this type of data allows to infer the activity cycles of Barcelona's population as well as the spatio-temporal distribution of their displacements. There are clear patterns of user behavior by station and type of day. From the temporal clustering of stations or by visualizing their average daily variation in activity it can be observed that the stations with similar behavior also correspond to adjacent areas in the map showing residential, university and leisure areas. The cycles allow a prediction of the amount of available bicycles in the stations, which is significantly better for time windows greater than 20 minutes than the current approach on the Bicing website where only the actual number of bicycles/free slots is shown. It is our intention to provide a prototype for an improved Bicing web interface which will allow such detailed predictions in the near future.

It would be interesting to contrast our results with more specific usage statistics. The Bicing system must internally produce more information that is not public, such as the origin/destination of individual users. Access to this data would allow to better validate our algorithm for the detection of the most probable routes and produce more precise models. Other information is not even available to the Bicing operator: e.g. the users that could not take/leave a bicycle because the station was empty/full. A survey aimed at obtaining a more detailed picture of the Bicing users and their motivations, currently being carried out by Jon Froehlich et al.[6], could help uncover this information.

A growing number of community bicycle services are appearing world wide[7], some of them with a similar web-service as the one we used to obtain our data, which is sure to generate increasing interest in this research topic.

[6] https://catalysttools.washington.edu/webq/survey/jfroehli/56481

[7] A huge list of such services can be found at the Bike-sharing world map at http://bike-sharing.blogspot.com.